# Attosecond charge transfer in atomic-resolution scanning tunnelling microscopy

S. Maier[1], R. Spachtholz[1], K. Glöckl[1], C. M. Bustamante[2], S. Lingl[1], M. Maczejka[1], J. Schön[1], F. J. Giessibl[1], F. P. Bonafé[2], M. A. Huber[1], A. Rubio[2,3], J. Repp[1], and R. Huber[1]

[1]Department of Physics and Regensburg Center for Ultrafast Nanoscopy, University of Regensburg, 93040 Regensburg, Germany

[2]Max Planck Institute for the Structure and Dynamics of Matter, 22761 Hamburg, Germany

[3]Initiative for Computational Catalysis, Flatiron Institute, 162 5th Avenue, New York, NY 10010, USA

**Electrons in atoms and molecules move on attosecond time scales. Deciphering their quantum dynamics in space and time calls for high-resolution microscopy at this speed[1,2]. While scanning tunnelling microscopy (STM) driven with terahertz pulses has visualized sub-picosecond motion of single atoms[3-15], the advent of attosecond light pulses has provided access to the much faster electron dynamics[16-19]. Yet, combining direct atomic spatial and attosecond temporal resolution remained challenging. Here, we reveal atomic-scale quantum motion of single electrons in attosecond lightwave-driven STM. Near-infrared single-cycle waveforms from phase-controlled optical pulse synthesis steer and clock electron tunnelling. By keeping the thermal load of the tip-sample junction stable, thereby eliminating thermal artifacts, we detect waveform-dependent currents on sub-cycle time scales. Our joint theory-experiment campaign shows that single-cycle near-infrared pulses can drive isolated electronic wave packets shorter than 1 fs. The angstrom-scale decay of the tunnelling current earmarks a fascinating interplay of multi-photon and field-driven dynamics. By balancing these effects, we sharply image a single copper adatom on a silver surface with lightwave-driven currents. This long-awaited fusion of attosecond science with atomic-scale STM makes elementary dynamics of electrons inside atoms, molecules and solids accessible to direct spatio-temporal videography and atom-scale petahertz electronics[7,20].**

**Introduction**

In an ongoing revolution, ultrafast microscopy has changed our perception of elementary dynamics on the nanoscale[1-15,21-27]. Lightwave-driven STM in the terahertz (THz) spectral range (THz-STM) for instance, has enabled atomic-scale slow-motion imaging of molecular orbitals and defect states with intrinsic spatio-temporal resolution[3-6,8-15]. Deep in the strong-field regime of light-matter-interaction[28], the oscillating carrier field has steered tunnelling across the tip-sample junction in a quasi-static manner[22,23]. Electrons have tunnelled only during the field crest of an oscillation half-cycle defining a time window as short as ~100 fs (ref. 3,4,8). By scaling the carrier frequency to the mid-infrared, a time resolution of 30 fs has been reached[24]. A direct observation and control of single-electron motion in the electronvolt energy range calls for single-digit-femtosecond or even attosecond resolution[29].

Meanwhile, petahertz ($10^{15}$ Hz) control of electrons has become a reality[7,20]. Near-infrared (NIR) light pulses centred at a frequency of 200 THz have driven sub-fs charge transfer in nanostructures[30-33] or attosecond field emission from metallic nanotips[34-38]. Yet the combination of attosecond time resolution and true atomic spatial resolution in STM has remained an ambitious goal. Simply scaling the carrier wave frequency in lightwave-driven STM by two orders of magnitude poses fundamental challenges, as multi-photon absorption processes become increasingly likely[28,39,40], which potentially compromises the sub-cycle control of field-driven tunnelling.

Here, we demonstrate sub-fs lightwave-driven STM with atomic resolution. Using a two-colour pulse synthesis[41] of phase-controlled near-infrared single-cycle waveforms allows us to keep a stable thermal load while resolving field-driven electron dynamics. Our approach eliminates thermal artifacts that would be dominant otherwise. In a joint theory-experiment approach, we identify the sweet spot of multi-photon and field-driven dynamics for optimal spatio-temporal resolution, demonstrated by imaging a single copper adatom on a silver surface with attosecond currents. Attosecond videography of atoms and molecules can now address diabatic electron dynamics and, in combination with atom manipulation, may promote the development of atom-scale petahertz electronics[7,20].

**Operating STM with NIR pulses**

We focus two spectrally non-overlapping time-delayed NIR pulses onto the tip-sample junction of a low-temperature STM and record the light-induced tunnelling current (Fig. 1a). Our experiments explore the crossover between two prominent limits of light-matter interaction (Fig. 1b) classified by the Keldysh parameter, $\kappa \propto \omega\sqrt{\epsilon_i}/E_0$, with the ionization potential $\epsilon_i$, the driving frequency $\omega$ and the electric field strength $E_0$ (ref. 28). In the (multi-)photon regime ($\kappa \gg 1$, upper panel), weak fields at high frequencies drive photon-assisted tunnelling via real or virtual states that experience a reduced tunnelling barrier[39]. In the strong-field regime ($\kappa \ll 1$, lower panel), slowly oscillating atomically strong fields act as an instantaneous bias that tilts the potential barrier, facilitating quasi-adiabatic tunnelling at the Fermi level, as used in THz-STM[3-15,22] and NOTE microscopy[23]. Improving the time resolution by tuning the carrier wave and the envelope of the field transients to shorter time scales inevitably pushes light-matter interaction towards the multi-photon regime, where sub-cycle tunnelling is challenged. To maintain strong-field interaction, we need single-cycle near-infrared pulses (~200 THz) with proportionally increased amplitudes of tens of MV/cm in the tunnelling junction.

Scaling up frequency and field strength by orders of magnitude strongly affects the thermal load on the tip-sample junction. We quantify the effect of laser power modulations on the junction stability by focusing the same train of femtosecond laser pulses (repetition rate, 80 MHz; centre wavelength, 1560 nm) onto the junction as in the primary experiments (see below). Slight variations of the pulse energy (mean value, 106 pJ) with an acousto-optic modulator (frequency, 874 Hz) introduces prominent modifications of the tunnelling current, which can be identified as thermally induced modulations $I_{th}$ of the total direct current $I_{DC}$. $I_{th}$ scales linearly with the power modulation depth (Fig. 1c, STM feedback loop with $I_{DC} = 100$ pA at $V_B = 200$ mV), indicating that the thermal current modulation $I_{th}/I_{DC}$ even exceeds the relative power modulation $\Delta P/P_0$. Most critically, for finite $V_B$, $I_{DC}$ is much larger than the coveted lightwave-driven current (see below). Because $I_{th}$ is synchronized with power modulations it is easily confused with lightwave-driven currents. This highlights the need of extraordinary stability of the relative power better than $10^{-4}$, precluding beam chopping for signal retrieval.

**Waveform synthesis for attosecond current control**

We employ a newly developed laser system that allows us to rapidly modulate the carrier-envelope phase (CEP), $\varphi_{CE}$, of the optical waveforms while keeping their power constant. The lightwave-driven current dynamics are explored with a two-pulse experiment with spectrally non-overlapping NIR pulses, which cannot interfere, excluding average power modulations. The CEP-stable pulses are generated in a custom-built Er:fiber laser system (TOPTICA). After amplification and broadening in highly nonlinear optical fibres, the spectrum spans more than one optical octave[41]. The spectral intensity and phase after pulse compression define two well separated pulses (Fig. 1d): the low- (purple, $\nu_c = 164$ THz) and high-frequency parts (orange, $\nu_c = 249$ THz) of the same continuum. Both are focused onto the junction of a low-temperature STM using high-numerical-aperture reflective optics. The mutual delay $\tau$ can be tuned with attosecond precision.

From the spectra of Fig. 1d, we can reconstruct the temporal shape of the field by an inverse Fourier transform up to a constant CEP offset. The intensity map of Fig. 1e shows the envelope of the field $E_{\text{env}}$ as a function of time $t$ and $\tau$, illustrating how the superposition of the two pulses modulates $E_{\text{env}}$. A line cut at $\tau = 0$ fs (Fig. 1f) traces a waveform with a width (FWHM) of the intensity envelope of 5.2 fs, corresponding to a single optical cycle at a centre frequency of 190 THz. In the diffraction-limited focus at the STM tip, peak electric fields of up to 7.6 MV/cm are reached. The inset of Fig. 1e illustrates two exemplary waveforms with $\varphi_{\text{CE}} = 0$ (red) and $\varphi_{\text{CE}} = \pi$ (dark red). The former exhibits a 1.36 ratio between the positive and negative field maxima. An acousto-optic phase shifter continuously modulates the CEP linearly in time as $\varphi_{\text{CE}}(t) \propto 2\pi f_{\text{CEO}} t$, where $f_{\text{CEO}} \approx 1$ kHz. The fraction of the tunnelling current that depends on the CEP is thus modulated at $f_{\text{CEO}}$. This component, called $I_{\text{CEO}}$, is detected by lock-in demodulation, yielding also its phase $\phi_{\text{CEO}}(\tau)$ in relation to $\varphi_{\text{CE}}(t)$.

As $\tau$ varies, the field envelope $E_{\text{env}}(t)$ alternates between single- and multi-cycle waveforms (Fig. 1e). Accordingly, the maximum of the field envelope $E_{\text{env}}^{\text{max}}$ oscillates with $\tau$ (Fig. 1g) and the carrier phase at the pulse maximum, $\phi_{\text{max}}$ exhibits a dominantly linear dependence on $\tau$ (Fig. 1h). Beside $\varphi_{\text{CE}}$, $\tau$ thus represents a second independent knob to change the waveform and the peak field without changing the intensity, and it grants access to sub-cycle tunnelling dynamics[37]. The indispensable power stability of the synthesized pulse train was monitored with a photodiode. While the CEP modulation scheme is active, the measured average power $P(\tau)$ (Fig. 1i, black curve) remains constant. The demodulated component of the power $P_{\text{CEO}}$ at $f_{\text{CEO}}$ (grey curve) reveals a minute relative modulation of $2 \times 10^{-5}$, which however does not depend on $\tau$.

**Fingerprints of sub-cycle tunnelling**

After rigorously excluding artefactual contributions, we search for genuine lightwave-driven tunnelling on an atomically flat Ag(100) surface (Fig. 2a, inset, red dot, $V_B = 200$ mV, $I_{set} = 100$ pA). Importantly, we find a stable and highly reproducible CEP-dependent component of the tunnelling current proving the existence of sub-cycle charge transfer. Figure 2a depicts the measured amplitude $I_{CEO}(\tau)$ for four pulse energies $\mathcal{E}_p = $ 171 pJ, 93 pJ, 55 pJ, and 36 pJ. The four curves are qualitatively similar but scaled in amplitude. $I_{CEO}(\tau)$ occurs only in the vicinity of $\tau = 0$, where the two pulses overlap to form the most asymmetric waveforms. Most remarkably, $I_{CEO}(\tau)$ exhibits strong oscillations on a sub-cycle scale. Reproducible time structures occur even on sub-fs scales (inset). The scaling of the maxima of $I_{CEO}(\tau)$ with $\mathcal{E}_p$ (Fig. 2b) follows a threshold-like behaviour, which underpins the nonlinear character of the tunnelling process.

$I_{CEO}(\tau)$ comes with a well-defined phase $\phi_{CEO}(\tau)$ (Fig. 2c, blue curve) with a dominantly linear dependence on $\tau$ like $\phi_{max}(\tau)$ (Fig. 2c, black curve, and Fig. 1h). Subtracting the linear dependence reveals regularly spaced steps (Fig. 2d), reminiscent of those in $\phi_{max}$, just with a smaller amplitude. This corroborates that $I_{CEO}$ is phase-locked to the absolute CEP, and therefore susceptible to the exact waveform – a hallmark of the strong-field regime. A light-induced tunnelling current that is CEP modulated and critically depends on the maximum electric field and the synthesized waveform clearly attests to sub-cycle charge transfer on attosecond time scales.

**Full quantum theory of NIR-induced tunnelling**

To microscopically understand this process, we strive for a full time-domain quantum theory including strong-field and multiphoton physics as well as their CEP dependence[42]. We simulate the light-induced charge transfer between two atomic sodium clusters representing the tunnelling junction with time dependent density functional theory (TD-DFT). The distance of the two clusters is 16 Å. In analogy to the experiment, the junction is biased with single-cycle field transients (peak amplitude, 1.04 V/nm) generated by the superposition of two pulses centred at 243 THz and 174 THz. The exemplary field transient in Fig. 3a ($\tau = -0.6$ fs, $\varphi_{CE} = 0$) drives a transfer of charge $Q(t)$ (Fig. 3b, dark red) that sharply sets on around $t = 0$ fs indicating a sub-half-cycle electron transfer event from the tip to the sample. The exact shape of $Q(t)$ and the total amount of the transferred charge at late times strongly depends on the CEP as demonstrated for $\varphi_{CE} = 0, \pi/2, \pi$ (colour coded from dark to light red). The transferred charge per laser shot $Q(t \to \infty, \varphi_{CE})$ multiplied with the repetition rate corresponds to a current. The theoretically retrieved CEP-dependent modulation depth $I_{CEO}^{DFT}(\tau)$ derived from $Q(t \to \infty, \varphi_{CE})$ and its phase $\phi_{CEO}^{DFT}(\tau)$ faithfully reproduce the experimentally measured $I_{CEO}(\tau)$ and $\phi_{CEO}(\tau)$, as shown in Fig. 3c-e for an exemplary pulse energy of $\mathcal{E}_p = 75$ pJ. The slope of $\phi_{CEO}^{DFT}(\tau)$ perfectly follows $\phi_{CEO}(\tau)$ (Fig. 3d). Based on the excellent theory-experiment agreement, we can use the TD-DFT results to scrutinize detailed features in the experimental data.

$I_{CEO}^{DFT}$ and $I_{CEO}$ strongly depend on $\tau$. The microscopic origin of this behaviour can be explained by comparing the waveforms for different delay times $\tau$ (Fig. 3f). Starting with a single-cycle transient ($\tau = -0.6$ fs, red curve), the carrier-wave maximum is shifted against the envelope maximum, for $\tau = -0.1$ fs (rose waveform), creating a CEP offset which gradually changes as $\tau$ increases. Simultaneously, an oscillation node shifts through the pulse, eventually destroying the single-cycle character. At $\tau = 0.9$ fs and 1.4 fs the envelope features two almost equally strong envelope maxima (purple dots). For 0.9 fs $< \tau <$ 1.4 fs, the maximum shifts from before the node to after the node, causing a step in $\phi_{max}(\tau)$.

This sudden CEP change consistently manifests as a step in both the experimental $\phi_{CEO}(\tau)$ and the computed $\phi_{CEO}^{DFT}$ (Fig. 3e), occurring within only a few hundred attoseconds in $\tau$. At these delay times, $I_{CEO}^{DFT}(\tau)$ is minimal, reflecting waveforms that feature several local minima and maxima of comparable height. The data points in Fig. 3c left of the minimum of $I_{CEO}^{DFT}(\tau)$ correspond to the first four waveforms. Starting from the most asymmetric single-cycle transient at $\tau = -0.6$ fs, increasing $\tau$ decreases the field asymmetry yielding a decrease in $I_{CEO}^{DFT}(\tau)$. In addition, the gradual change in CEP leads to the slope in $\phi_{CEO}^{DFT}(\tau)$ (first four data points in Fig. 3e). Thus, $I_{CEO}$ depends on subtle features of the waveform proving strict sub-cycle attosecond charge transfer.

With the full quantum-theory at hand we can further trace the current in space and time. Figure 3g displays four snapshots of the simulated tip-sample junction during the charge transfer process driven by the waveform in Fig. 3a ($\varphi_{CE} = 0$, grey vertical lines mark the corresponding times $t$). The difference in charge density $\Delta\rho(t, x, z)$ with respect to the unperturbed ground state $\rho_0$ is colour coded. At $t = -$ 24.4 fs, when the electric field is still rather low, Δρ reflects the geometry of tip (top) and sample (bottom). During the main negative ($t = -2.4$ fs) and positive ($t = -0.4$ fs) half cycles, a strong shake-up of $\Delta\rho(t)$ becomes apparent, with Δρ occupying increasingly more volume. At $t = 1.1$ fs, the wave packet bridges the gap as indicated by the finite Δρ between tip and sample. Figure 3b shows the corresponding $Q(t)$ and its derivative, i.e. the current (Fig. 3a, purple area). The current flows within a time window of slightly less than 1 fs, confirming the sub-cycle nature of the field-driven charge-transfer process.

Most remarkably, the current maximum occurs with a distinct retardation of 0.5 fs after the field maximum and persists up to about 1.5 fs. Unlike in THz-STM on the 10 to 100 fs scale, the current response in attosecond STM is no longer instantaneous as we start to reach the response time of electrons in solids. The retardation of the charge transfer to the field is characteristic of the intermediate light-matter interaction regime, $\kappa \sim 1$, and related to the Keldysh time[28]. Moreover, the increasing evanescent Δρ extending into the vacuum region seen in Fig. 3g hints towards the occupation of states energetically close to the vacuum level and, therefore, a transient population of highly excited states, confirmed by the simulated change in occupation (Fig. 3h). These excited electrons, with reduced tunnelling barriers and extended wavefunctions finally cross the barrier in the lightwave-driven tunnelling regime – very distinct from photon-driven field emission.

**Imaging a single Cu adatom with attosecond currents**

This novel mechanism of photon-assisted lightwave-driven tunnelling, bridging different regimes, calls for an assessment of its spatial resolution. We first measured the distance dependence of $I_{\mathrm{CEO}}$ (Fig. 4a) by setting τ to the maximum of $I_{\mathrm{CEO}}$ and retracting the tip from the sample starting from a defined setpoint ($I_{\mathrm{DC}} = 1$ nA at $V_{\mathrm{B}} = 200$ mV). For the highest pulse energy of 171 pJ, the tunnelling current decays by one order of magnitude within $l_{\mathrm{c}} = 8.7$ Å. This decay constant is distinctly larger than characteristic of steady-state tunnelling at the Fermi level ($l_{\mathrm{c}} = 1$ Å) and attests to the contribution of excited electrons with an effectively reduced barrier and more extended wave functions. The simulated $I_{\mathrm{CEO}}^{\mathrm{DFT}}(\Delta z)$ faithfully reproduces the experimental decay length, assuming that a pulse energy of $\mathcal{E}_{\mathrm{p}} = 171$ pJ corresponds to a peak field of 1.04 V/nm.

To benchmark the actual lateral resolution of $I_{\text{CEO}}$ we image a single copper adatom on Ag(100). At a pulse energy of $\mathcal{E}_p = 93\,\text{pJ}$ the tip was first approached above the centre of the atom to $I_{\text{set}} = 100\,\text{pA}$ at $V_B = 200\,\text{mV}$ and subsequently raster-scanned in constant-height mode. The simultaneously recorded maps of $I_{\text{DC}}$ and $I_{\text{CEO}}$ are displayed in Figs. 4b and c, respectively. Both signals clearly resolve the atom; line profiles confirm the spatial resolution (Fig. 4d). Notably, $I_{\text{CEO}}$ decreases above the adatom. The exact contrast mechanism is currently under investigation. We tentatively assign it to the local decrease in the work function of the sample above the adatom[43,44], potentially leading to a locally more symmetric tunnelling barrier that reduces the directionality of lightwave-driven tunnelling and thereby $I_{\text{CEO}}$. These results demonstrate atomic-resolution imaging with an attosecond lightwave-driven tunnelling current.

**Conclusion**

Our novel lightwave-driven STM combines attosecond temporal resolution with atomic-scale imaging. By using two-colour waveform synthesis, we achieve attosecond waveform control while avoiding thermal measurement artifacts. Measurements of the CEP- and waveform-dependent tunnelling currents reveal intricate modulations that serve as a fingerprint of the underlying processes. They confirm charge transfer on attosecond time scales, with a CEP-induced modulation of less than one electron per shot. Theory-experiment comparisons reveal the details of sub-cycle tunnelling dynamics of a non-equilibrium electron distribution through an effectively reduced barrier. In contrast to THz-STM, tunnelling does not occur instantaneously with the electric field making the inherent reaction time of electrons accessible. Imaging of a single Cu adatom demonstrates atomic resolution in this previously uncharted regime of light-matter interaction. The possibility to image individual atoms while resolving attosecond dynamics establishes our new concept as a powerful tool for fundamental studies of electron motion in time and space. It transforms our ability to study and control quantum phenomena in molecules, nanoscale devices, and correlated materials.

**Data Availability**

Source data are provided with this paper. All further data are available from the corresponding authors.

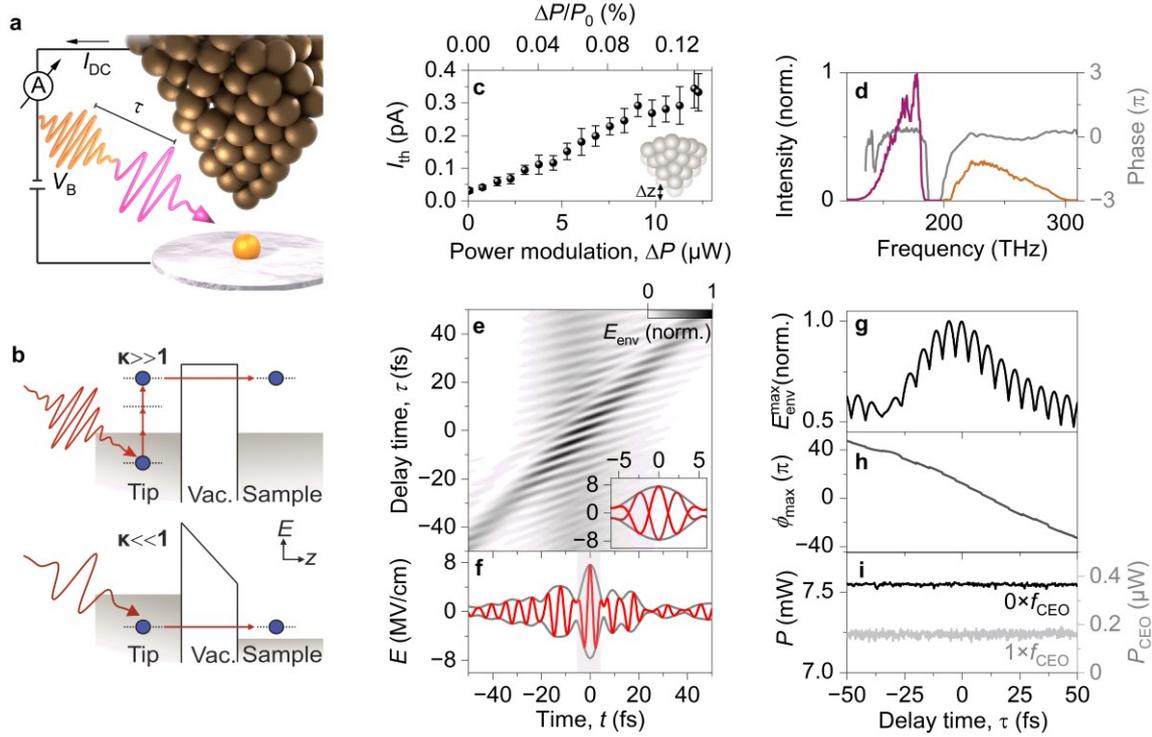

**Figure 1 | Attosecond lightwave-STM. a**, Schematic setup: Two spectrally distinct near-infrared (NIR) pulses with variable delay time τ are focused onto the sharp metal tip of a scanning tunnelling microscope. The bias voltage $V_B$ is applied to the sample with respect to the tip. The carrier-envelope phase (CEP) dependent part $I_{CEO}$ of the total tunnelling current $I_{DC}$ is extracted. **b**, Schematic representation of optically induced tunnelling in two limiting cases: For Keldysh parameters $\kappa \gg 1$, electrons (blue dot) can be excited via (multi-)photon absorption to tunnel (horizontal arrow) through the shallow remaining potential barrier (thick black line) with a strongly enhanced probability. In the strong-field limit ($\kappa \ll 1$), the oscillating electric field tilts the potential quasi adiabatically, and lightwave-driven tunnelling occurs in the energy window between the Fermi levels of tip and sample. **c**, Even microwatt-level modulations $\Delta P$ of the incident laser power $P_0 = 8.5$ mW (106 pJ) (relative modulation depth $\Delta P/P_0 \sim 10^{-4}$) lead to appreciable modulations $I_{th}$ of the DC current, owing to thermal effects (current set point, 100 pA at 200 mV bias). The offset is given by the noise floor (~50 fA). **d**, The intensity spectra of the two incident NIR pulses (soliton and dispersive wave) show no overlap, excluding spectral interference between them. The spectral phase (grey) is flat across significant portions of the spectrum. **e**, Envelope function $E_{env}$ of the superposition of the two ultrashort laser pulses as a function of delay time τ, reconstructed from the optical spectrum. The ultrafast modulation of the

envelope in time maximizes the dependence of the electric-field transients on the CEP, $\varphi_{CE}$ (inset). **f,** $E_{env}$ (grey) and resulting single-cycle electric field waveform (FWHM = 5.2 fs) at $\tau = 0$ and $\varphi_{CE} = 0$ featuring a pronounced field asymmetry with a peak electric field strength of up to 7.6 MV/cm in the far field. **g,** $E_{env}^{max}$ as a function of delay time $\tau$. **h,** $\varphi_{CE}$, at which the carrier field exhibits the largest peak as a function of $\tau$, $\phi_{max}$. **i,** The average laser power $P$ stays constant as a function of $\tau$ (left axis, black line). The power demodulated at $f_{CEO}$ shows a small amplitude of 160 nW which is independent of $\tau$ (grey, right axis, relative modulation $2\times10^{-5}$).

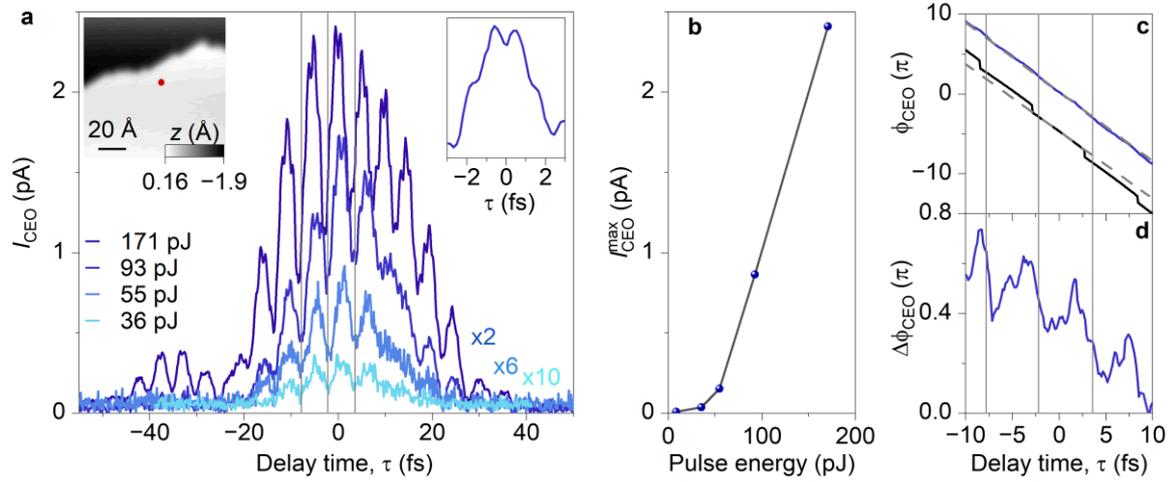

**Figure 2 | Evolution of lightwave-driven tunnelling currents with delay time $\tau$. a**, CEP–modulated tunnelling current $I_{CEO}$ as a function of $\tau$ for various total pulse energies $\mathcal{E}_p = $ 171 pJ, 93 pJ, 55 pJ, and 36 pJ (colour coded from dark to light blue, $I_{DC} = $ 100 pA, $V_B = $ 200 mV). For $\tau = 0$, the maxima of the envelopes of both NIR pulses coincide in time. Grey lines mark the positions of current minima for $\mathcal{E}_p = $ 93 pJ. Data sets for low pulse energies are multiplied by a constant factor as indicated. Right inset: Zoom-in of $I_{CEO}$ for $\mathcal{E}_p = $ 171 pJ. Left inset: Constant-current STM image of the sample surface. The tip position during the lightwave-current measurements is indicated with a red dot. After each measurement, the sample was re-inspected to exclude topographic changes, confirming nondestructive measurement conditions. **b**, Nonlinear scaling of the peak value of $I_{CEO}(\tau)$ with the pulse energy $\mathcal{E}_p$. **c**, Corresponding phase $\phi_{CEO}$ of $I_{CEO}$ (blue) for 93 pJ, compared to $\phi_{max}$ (black, c.f. Fig. 1h). Both curves exhibit a dominantly linear dependence with the same slope. Dashed lines: linear fits between two adjacent minima in $I_{CEO}(\tau)$ or $E_{env}^{max}(\tau)$, respectively. **d**, Phase difference $\Delta\phi_{CEO}$ after subtracting a linear fit from the experimentally measured phase $\phi_{CEO}$ (blue curve). $\Delta\phi_{CEO}$ exhibits steps at the positions of the current minima in **a** (grey vertical lines).

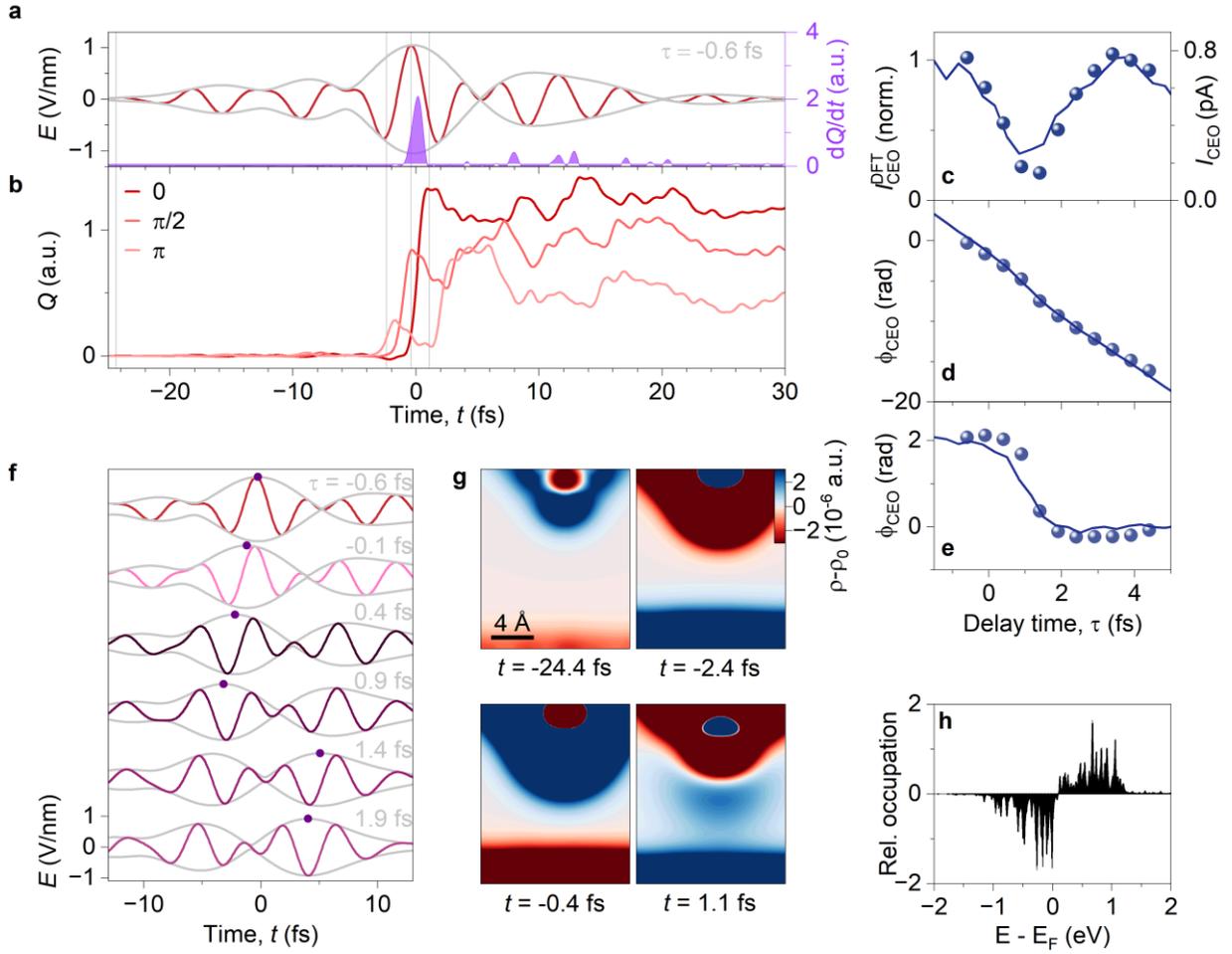

**Figure 3 | Time-domain DFT simulation of sub-cycle currents. a**, Electric field transient (red) and its envelope (grey) of an exemplary single-cycle transient (for $\tau = -0.6$ fs, $\varphi_{CE} = 0$ and peak field, 1.04 V/nm) used for the TD-DFT simulations of lightwave-driven currents between two atomic sodium clusters. The corresponding simulated current transient $I(t) = dQ/dt$ in purple is confined to a duration of 985 as (FWHM; $\varphi_{CE} = 0$). **b**, Simulated time-resolved charge transfer $Q$ between tip and sample for $\varphi_{CE} = 0, \frac{\pi}{2}, \pi$ (colour coded in red). The large step in $Q(t)$ near $t = 0$ fs, indicating charge transfer in atomic units (a.u.) from tip to sample, strongly depends on the CEP. **c**, CEP-modulated part of the transferred charge $I_{CEO}^{DFT}$ as a function of $\tau$ (dots) and an experimentally measured $I_{CEO}$ for $\mathcal{E}_p = 75$ pJ, $V_B = 0$ mV (setpoint, $I_{set} = 100$ pA at $V_B = 200$ mV). The measurement was taken above a Cu(111) surface. **d**, Phase of the simulated $I_{CEO}^{DFT}$ (dots) and corresponding experimental $\phi_{CEO}$. Both feature an overall linear behavior with local softened phase jumps. The theoretical data was shifted in $\tau$ and a constant offset was applied to match the experimental data. **e**, Phases from panel **d** with a linear slope ($\sim -\pi/\text{fs}$) subtracted. **f**, Electric field transients for different $\tau$. The overall linear behaviour in phase (**e**)

is consistent with the maximum of the envelope shifting in time with increasing τ, such that $\phi_{\text{CEO}}$, at which the maximum of the envelope coincides with a maximum of the carrier wave gradually decreases. For certain τ the maximum of the envelope (marked with a dot) exhibits sudden changes in time, explaining the phase jumps of $\phi_{\text{CEO}}$. These values of τ correspond to envelopes with low field asymmetry, and, therefore, a low $I_{\text{CEO}}^{\text{DFT}}$– as observed in the experiment. **g**, Selected snapshots of the spatial distribution of the simulated relative charge density $\Delta\rho(t) = \rho(t) - \rho_0$ (colour coded in atomic units) in the tunnel junction (tip-sample distance, 16 Å), with the equilibrium charge density $\rho_0$. The images represent a cut through the middle of the tip and the slab at $y = 0$. Emerging excitations at $t = $ -24.4 fs follow the geometry of the sodium clusters forming tip (top) and sample (bottom). At $t = $ -2.4 and -0.4 fs a strong shakeup of $\Delta\rho(t)$ becomes apparent, while $\Delta\rho$ inside the tunnelling gap indicates the rapid charge transfer at $t = 1.1$ fs. The colour scale saturates at $3 \times 10^{-6}$ a.u.. **h**, Simulated energy-resolved change in occupation of electronic states after an excitation of the tip-sample cluster with a gaussian pulse (peak electric field, 2.16 V/nm; $\nu_{\text{center}} = 242$ THz). States below (above) the Fermi energy $E_F$ get transiently depopulated (occupied).

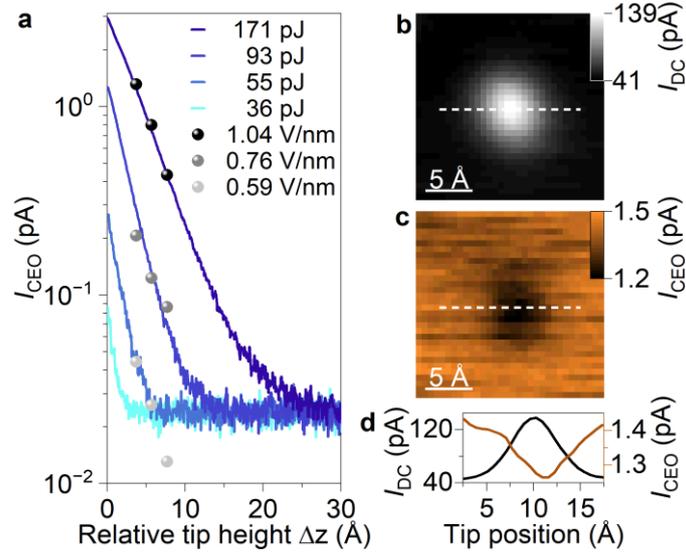

**Figure 4 | Atomically resolved sub-cycle currents. a**, $I_{CEO}$ as a function of tip-sample distance. At the highest pulse energy used $\mathcal{E}_p = 171$ pJ, the decay length of ~8.7 Å reflects tunnelling through a shallow remaining barrier. For pulse energies $\mathcal{E}_p < 100$ pJ, the decay becomes much steeper ($< 5$ Å). Exponential fit, $I \propto \exp(-2\kappa\Delta z)$, for $\mathcal{E}_p = 93$ pJ, corresponding to a decay constant of $\kappa = 0.2$ Å$^{-1}$. Black and grey dots are the simulation results for $I_{CEO}^{DFT}$. The data in **a** are averaged over 30 consecutive measurements. The offset is given by the measurement noise floor. **b**, Constant-height STM scan above a single Cu adatom on Ag(100). **c**, Simultaneously measured lightwave-induced tunnelling current for $\mathcal{E}_p = 93$ pJ demonstrating that attosecond tunnelling can sustain atomic-resolution imaging. **d**, Line-cut through **b** and **c** (dashed lines) proving similar lateral feature detail for DC and lightwave-induced currents.

**Acknowledgements** We thank T. Siday, L. Gross and Y. A. Gerasimenko for fruitful discussions and T. Buchner, H. Friedrich, K. Pürckhauer, A. Schüller, D. Riese, L. Sporrer, I. Gronwald, M. Furthmeier, M. Heinl, I. Laepple and C. Rohrer for their contributions to instrument development. The work in Regensburg has been supported by the Deutsche Forschungsgemeinschaft (DFG, German Research Foundation) through Project HU1598/8, RE2669/8, INST 89/414-1 FUGG, 314695032, 406658631, 502572516 and by the European Union through the European Research Council Synergy grant no. 951519, MolDAM. F.P.B. acknowledges financial support from the European Union's Horizon 2020 research and innovation program under the Marie Sklodowska-Curie Grant Agreement No. 895747 (NanoLightQD). C.M.B. thanks the Alexander von Humboldt-Stiftung for the financial support from the Humboldt Research Fellowship. C.M.B. and F.P.B. also would like to acknowledge the computational support provided by the Max Planck Computing and Data Facility. This work was supported by the European Research Council (ERC-2024-SyG- 101167294 ; UnMySt), the Cluster of Excellence Advanced Imaging of Matter (AIM), Grupos Consolidados (IT1249-19) and SFB925. We acknowledge support from the Max Planck-New York City Center for Non-Equilibrium Quantum Phenomena. The Flatiron Institute is a division of the Simons Foundation.

**Author Contributions** S.M., F.J.G., J.R. and R.H. conceived the study. J.R. and R.H. supervised the study. S.M., R.S., K.G., S.L., M.M., J.S., M.A.H., J.R. and R.H. carried out the experiment, analysed the data and contributed to the discussions of the experimental results. C.M.B., F.P.B. and A.R. developed and performed the TD-DFT computations and analysed the data together with S.M. and R.H.. S.M., R.S., K.G., J.R. and R.H. wrote the manuscript with contributions from all authors.

**Competing interests.** The authors declare no competing interests.

**Correspondence.** Correspondence and requests for materials should be addressed to J.R. (jascha.repp@ur.de) or R.H. (rupert.huber@ur.de).